\documentclass[11pt,a4paper]{article}
\usepackage{jheppub}
\usepackage[utf8]{inputenc}
\usepackage{physics}
\usepackage{feynmp-auto}
\usepackage{array}
\newcolumntype{P}[1]{>{\centering\arraybackslash}p{#1}}
\usepackage{mathrsfs}
\usepackage{graphicx}
\usepackage{natbib}
\usepackage{caption}
\usepackage{makecell}
\usepackage{tikz}
\usepackage{subfig}
\makeatother
\usepackage{multirow}
\newcommand{\marrow}[5]{%
    \fmfcmd{style_def marrow#1
    expr p = drawarrow subpath (1/4, 3/4) of p shifted 12 #2 withpen pencircle scaled 0.4;
    label.#3(btex #4 etex, point 0.35 of p shifted 18 #2);
    enddef;}
    \fmf{marrow#1,tension=0}{#5}}
\newcommand{\marrows}[5]{%
    \fmfcmd{style_def marrows#1
    expr p = drawarrow subpath (1/4, 3/4) of p shifted 0 #2 withpen pencircle scaled 0.4;
    label.#3(btex #4 etex, point 0.5 of p shifted 6 #2);
    enddef;}
    \fmf{marrows#1,tension=0}{#5}}
    
\newcommand{\marrowss}[5]{%
    \fmfcmd{style_def marrowss#1
    expr p = drawarrow subpath (1/2, 7/8) of p shifted 3 #2 withpen pencircle scaled 0.4;
    label.#3(btex #4 etex, point 0.75 of p shifted 5 #2);
    enddef;}
    \fmf{marrowss#1,tension=0}{#5}}
\usepackage{fancyhdr}
\usepackage{longtable}
\usepackage{parskip}
\usepackage[T1]{fontenc}
\usepackage{dcolumn}   

\usepackage{bm}        
\usepackage{amsfonts}  
\usepackage{amsmath}   
\usepackage{amssymb}


\usepackage{natbib}

\preprint{CERN-TH-2020-148}

\title{Charting the Fifth Force Landscape}
\author[a]{Hannah Banks}
\author[b,a]{and Matthew McCullough}
\affiliation[a]{DAMTP, University of Cambridge, Wilberforce Road, Cambridge, UK}
\affiliation[b]{CERN, Theoretical Physics Department, Geneva, Switzerland}
\abstract{Fifth forces arising from hidden sector scalar operator exchange, as probed in a variety of present and planned experiments, can be captured by a general dispersion relation involving one real, positive, spectral density function.  Previously considered scalar fifth forces, from tree-level to more exotic loop-level possibilities, are derived in this formalism, without explicitly performing conventional loop calculations.  A variety of experimental observables commonly used in fifth force searches are also presented in the same formalism, allowing the straightforward extraction of limits on any specific model.  The speculative possibility of probing hidden sector violations of unitarity, causality, or locality, manifested as a breakdown of the positivity of the spectral density, is also discussed.}
\begin{document}
\maketitle

\section{Introduction} 
It has taken decades to explore the dark side of the Standard Model (SM), and there is still much left to uncover. For example, although they harbour just a fraction of the cosmological SM energy quota, neutrinos provide a unique window through which to study the Universe. Similarly, light states from dark sectors may also have much to tell us. The exchange of new light particles by ordinary nucleons would generate an additional, and potentially observable,
`fifth' force between them. A plethora of experimental searches have been initiated in this vein, looking for new weakly coupled dark sector states. With current and near-future advances in quantum sensing technologies unlocking a host of new search avenues at the precision frontier (see, for example, Refs.~\cite{Safronova_2018,Biedermann:2014jya,Jones_2020,Berengut_2018}), the hunt has entered a new and exciting chapter. 

Any search for new states requires direction, otherwise experimental effort may be spent in unpromising phenomenological corners.  Often that direction comes from considering toy theoretical models which are simple enough to provide concrete predictions whilst, one hopes, still capturing the relevant phenomenological characteristics of more complex, realistic, constructions. Interpretations of fifth force experiments to-date have primarily focused on the simplest such model: tree-level exchange, although some theoretical attention has been given to more exotic possibilities. In Ref.~\cite{Ferrer_1998}, spin-independent forces arising from double pseudo-scalar exchange were considered using a dispersion technique \cite{FeinbergG1989Tdto}. More recently, the spin-independent inter-nucleon potentials generated by the exchange of a sample of scalar and vector composite operators comprising of pairs of dark scalars, fermions and vectors, have been calculated within an effective field theory framework \cite{Fichet_2018}. Experimental constraints on those involving scalar fields were determined in Ref.~\cite{Brax_2018} using the results from a number of recent fifth force experiments. Some of the potentials and  exclusion limits computed in these works will be re-examined in this paper through a new lens.  A number of possibilities for spin-dependent forces arising from various theoretical scenarios have also been presented in Refs.  \cite{Dobrescu:2006au,Fadeev:2018rfl,Costantino_2020}.
The variety in force properties revealed in these examples motivates a broader study of the fifth force landscape, which to-date remains scantly explored. Indeed, very recently it was argued that new quantum forces may be linked to some muonic puzzles \cite{Perelstein:2020suc}, providing further motivation for a systematic exploration of theoretical possibilities.

Ideally, to escape the inherent limitations of model-by-model investigations, one seeks a completely general description that  would encapsulate the broadest possible set of features.  While this appears, from a theoretical perspective, too cumbersome to be useful, there are some scenarios in which the basic properties of quantum field theories (QFTs) are sufficient to delineate a broad swatch of possibilities, without needing to resort to a significant number of toy models.

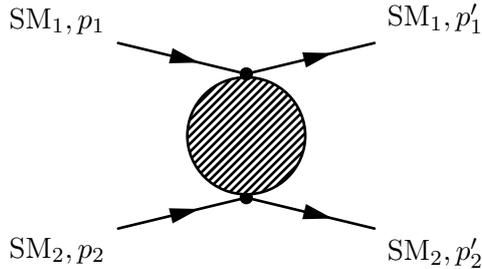
\begin{figure}[t]
    \centering
\begin{fmffile}{diagram}
\begin{fmfgraph*}(120,70)
   \fmfleftn{i}{2}
    \fmflabel{$\text{SM}_2,p_2$}{i1}
    \fmflabel{$\text{SM}_1,p_1$}{i2}
   \fmfrightn{o}{2}
   \fmflabel{$\text{SM}_2,p_2'$}{o1}
    \fmflabel{$\text{SM}_1,p_1'$}{o2}
     \fmf{fermion}{i1,v1,o1}
     \fmf{phantom}{v1,v2,v3}
  \fmfblob{44}{v2} 
\fmfdot{v1}
\fmfdot{v3}
\fmf{fermion}{i2,v3,o2}
\end{fmfgraph*}
\end{fmffile}
\vspace*{10mm}
    \caption{Factorisation of $i\mathcal{M}$ for SM fermion scattering into the  external SM states and the mediating hidden sector ones in $\Delta(q)$, represented by the shaded circle.  }
    \label{fig:generalised}
\end{figure}

Consider the case where some SM states, contained within the gauge-invariant scalar composite operator $\mathcal{O}_{SM}$, are very weakly coupled to some new dark sector states within the scalar composite operator $\mathcal{O}_{DS}$. If sufficiently light, these states may generate miniscule forces in terrestrial experiments.  By `force' we mean a process in which both the initial and final states contain only SM particles i.e.\ $\text{SM}\to\text{SM}$ scattering. We assume that the dark sector states are not cosmologically abundant and that scattering of the form $\text{SM}+\text{DS}\to\text{SM}+\text{DS}$ is not phenomenologically relevant.  We may write such an interaction as
\begin{equation}
\mathcal{L}_{\text{int}} = c~ \mathcal{O}_{SM} \mathcal{O}_{DS} ~~,
\end{equation}
where $c$ is a small coupling.  At $\mathcal{O}(c)$, the only corrections to SM physics are through the renormalisation of SM operators, which can be absorbed into their parametric definition.  At $\mathcal{O}(c^2)$ however, as depicted in Fig.~\ref{fig:generalised}, new forces arise whenever the dark sector states (whatever they may be) are integrated out.

Since we have not specified any details of the dark sector, this scenario seems too general to be predictive.  The goal of this work is to show that this is not the case.  The reason is that the nature of the force is determined by the correlation function
\begin{equation}
\langle \mathcal{O}_{DS} (x) \mathcal{O}_{DS} (y) \rangle ~~,
\end{equation}
which, thanks to insightful work in the early days of quantum field theory, admits a general representation in terms of a single integral: the K\"all\'en-Lehman representation.  As a result, any new scalar force arising from the exchange of dark sector states may be characterised in a simple, minimal, manner. In this work we exploit this to propose a new approach to parameterising dark sector forces, underpinned by the K\"all\'en-Lehman representation, as a unified, general, language for organising the exploration of dark fifth forces.  

To demonstrate the power of this approach, after deriving the main results in Sec.~\ref{GP}, we show in Sec.~\ref{ESD} that many of the predictions of toy models previously considered in the literature concerning weakly coupled fifth forces, from tree-level to loop-level particle exchange, may be obtained without resorting to conventional loop calculations.  Such models apply to dark sectors whose internal dynamics can be treated perturbatively.  It is equally plausible, however, that dark sectors may involve strongly coupled dynamics.  In this case, perturbative models will fail and only general non-perturbative tools such as the K\"all\'en-Lehman representation can be trusted to faithfully describe the underlying processes.  In Sec.~\ref{landscape} we map a sample of interesting possibilities for dark sector-mediated forces by calculating present experimental bounds on a handful of example scenarios.  Importantly, the relevant experimental observables for spin-independent fifth force searches, presented in Sec.~\ref{ES}, are expressed in general terms, such that bounds may be cast for any dark sector fifth forces with ease.

We may also use this formalism to look beyond QFT.  It is possible that some concepts fundamental to quantum field theory such as locality, unitarity, or causality may break down at ultra-microscopic scales in the deep UV, perhaps related in some way to the UV-completion of quantum gravity.  However it is also possible, but perhaps not so plausible, that these fundamentals may be violated in the IR at scales accessible to experiment, having been hidden thus far due to an extremely weak coupling rather than an inaccessibly high energy scale.  Were this the case, one would wish to be able to identify not only the presence of some new sector of states beyond the SM, but also whether there are any tell-tale signs of the breakdown of these fundamentals in experimental observables.  To this end we note that the K\"all\'en-Lehman representation has these properties hardwired into one feature: positivity of the integrand. This leads to forces that are a monotonic function of distance. The general framework proposed here thus allows, at least in principle, to diagnose when some fundamental property of QFT is potentially being violated.  However, this is only under some important assumptions.  The first is the assumption of scalar operator exchange.  Since different spin operators can generate forces of different signs, a non-monotonic force profile could also arise from the interference of different spin operators.  If it could be experimentally determined that the operator is scalar, then this assumption would not be necessary.  A second is the assumption that the hidden sector states are all positive-norm.  This may seem a relatively safe assumption, however hidden sector gauge groups could, in principle, contribute negatively through intermediate negative norm states \cite{Cacciapaglia:2006pk}.  Finally, we emphasise that even under the assumption of scalar operator exchange, breakdown of positivity may not reveal precisely \emph{which} fundamental concept is breaking down.  Nonetheless, with these caveats in mind, by employing the simple mapping from the general K\"all\'en-Lehman representation to experimental observables presented in Sec.~\ref{ES}, we explore in Sec.~\ref{landscape} some potential experimental features that may reveal not only the existence of a hidden sector, but also potentially place interesting experimental limits on the presence of non-trivial dynamics which could signify a breakdown of QFT itself within the dark sector.

\section{From QFT Amplitudes to Potentials}\label{GP}
We first review how particle exchange generates a potential within the framework of QFT. Consider the scattering of two distinguishable fermions $\Psi$, of equal mass, $M$, and initial 4-momenta $p_{1}$ and $p_{2}$ to states of 4-momenta $p'_{1}$ and $p'_{2}$ respectively. We denote the QFT amplitude for this process to be $i\mathcal{M}$. To extract the force felt by each fermion, we turn to the description of scattering in non-relativistic quantum mechanics. In Born's approximation, the leading order amplitude  $A_{\textnormal{NR}}$ for a particle of 3-momentum $\boldsymbol{p}$ to scatter in a potential $V(r)$ to a momentum $\boldsymbol{p'}$ is
\begin{equation}
    \label{ANR}
    A_{\textnormal{NR}} = -i \int d^3 \boldsymbol{r} ~ e^{-i((\boldsymbol{p}'-\boldsymbol{p}) \vdot \boldsymbol{r})} V(\boldsymbol{r}) ~~.
\end{equation}
 The amplitude $i\mathcal{M}$ can be compared with $A_{NR}$ on taking the non-relativistic limit, and dividing by a factor of $4M^2$ to account for the implicit use of relativistically normalised states. The inter-fermion potential can thus be identified as  
\begin{equation}
\label{pot}
    V(\boldsymbol{r}) = - \frac{1}{4M^2}\int d^{3} \boldsymbol{q}~ \frac{\mathcal{M}^{\textnormal{NR}}}{(2\pi)^3} e^{i \boldsymbol{q}\vdot \boldsymbol{r}} ~~,
\end{equation}
where $\boldsymbol{q} = \boldsymbol{p}'-\boldsymbol{p} $ is the momentum transfer between the fermions, and the superscript NR indicates evaluation in the non-relativistic limit.  
\subsection{The Yukawa Potential}
Consider the case where the fermion interaction is mediated by the tree-level exchange of a single scalar, $\phi$, of mass $m$ and Yukawa coupling $\lambda$. In the non relativistic limit, $ M  \gg \boldsymbol{p}$, the amplitude becomes 
\begin{equation}\label{amp}
    i\mathcal{M}^{\textnormal{NR}} = -4i\lambda^2  M^2 \delta^{s_1 s_1'}\delta^{s_2 s_2'}\Delta_F^{\textnormal{NR}} ~~,
\end{equation}
where \begin{equation}
\Delta_F^{NR} = - \frac{1}{m^2+ | \boldsymbol{q}|^2} ~~,
\end{equation}
is the (non-relativistic) propagator for $\phi$ and $s_{1,2}$ ($s'_{1,2}$) denote the spin polarisations of each of the incoming (outgoing) fermions.  Inserting this into Eq.~\ref{pot}, converting to polar co-ordinates and using a contour integral to evaluate the integral over $|\boldsymbol{q}|$ in the complex plane yields the familiar Yukawa potential: \begin{equation}
    V(r) = - \frac{\lambda^{2}}{4\pi}\frac{e^{-mr}}{r} ~~.
\end{equation}
We now employ this textbook result to generalise beyond Yukawa forces.

\subsection{Generalised Potentials}
We seek to develop a prescription capable of describing the potential generated by \textit{any} possible scalar operator exchange. Assuming that the hidden sector couples only weakly to the SM, the amplitude for SM scattering will also take the form of Eq.~\ref{amp} for some $\Delta(q)$ which encapsulates the mediating dark sector dynamics. The factorisation of the amplitude into the external SM, and internal hidden sector states in this way is shown schematically in Fig.~\ref{fig:generalised}.

In direct analogy with the Yukawa interaction, we can think of this as being a tree-level exchange generated from an effective $\sim \lambda\bar{\Psi}\Psi\Phi$ term in the Lagrangian, for some scalar operator $\Phi$, which is itself a function of the elementary hidden sector fields. In this picture, $\Delta$ plays the role of an effective scalar propagator for $\Phi$. As with any scalar 2-point function, it is therefore possible to express $\Delta$ in  K\"all\'en-Lehmann spectral form  as
 \begin{equation}
 \label{KL}
\Delta(q) = 2 \int_0^{\infty} \mu d\mu ~ \frac{\rho(\mu^2)}{q^2 - \mu^2 + i\epsilon} ~~,
\end{equation}
for some  real and positive-definite spectral density $\rho(\mu^2)$ \cite{Weinberg:1995mt}. Inserting the non-relativistic limit of Eq.~\ref{KL} into Eq.~\ref{pot} yields 
 \begin{equation}
 \label{potential}
     V(r)  = -\frac{\lambda^2}{2\pi r}\int_0^{\infty} \mu d\mu ~ \rho(\mu^2)e^{-\mu r} ~~.
 \end{equation}
 This is the most general form of the potential generated from scalar operator exchange allowed within the general postulates of QFT.
 
\section{Extracting the Spectral Density}\label{ESD}

To find $\rho(\mu^2)$ for a specific interaction, we note that 
\begin{equation}
\label{eq:rho}
    \rho(q^2) = -\frac{1}{\pi}\Im{\Delta(q)} ~~.
\end{equation}
It is clear that for tree-level scalar exchange, $\rho(\mu^2) = \delta(\mu^2 - m^2)$. Indeed, inserting this into Eq.~\ref{potential} recovers the Yukawa potential. As we shall see, computing the scattering potential directly from $\rho(\mu^2)$ offers considerable advantage in the case of loop exchange, since only the imaginary part of $i\mathcal{M}$ is required, bypassing the need for conventional loop calculations.

\subsection{One-Loop Exchange}
To illustrate the utility of this spectral representation to describe scalar mediated scattering, we first consider a set of specific toy hidden sector models, each comprised of a single dark field coupled bilinearly to SM fermions. Denoting a generic spin 0, spin 1/2 and spin 1 hidden field of mass $m$  by $\phi$, $\psi$ and  $V^{\mu}$ respectively, we consider, in turn, the following extensions to the SM Lagrangian \begin{equation}
\label{boxed}
\textnormal{(A)} \hspace{0.5em}\frac{1}{\Lambda} \mathcal{O}_{SM}|\phi|^2 \hspace{1.5em} \textnormal{(B)} \hspace{0.5em} \frac{1}{\Lambda^{2}} \mathcal{O}_{SM}\Bar{\psi}\psi \hspace{1.5em} \textnormal{(C)} \hspace{0.5em} \frac{m^2}{\Lambda^3} \mathcal{O}_{SM}|V|^2\hspace{1.5em} \textnormal{(D)} \hspace{0.5em} \frac{1}{\Lambda^3} \mathcal{O}_{SM}\partial_{\mu}\phi^{*}\partial^{\mu}\phi
 ~~,\end{equation}
where $\mathcal{O}_{SM}$ is the bilinear SM fermionic operator  $\bar{\Psi}\Psi$  evaluated below the QCD scale. For practical purposes this will be a bilinear of nucleons, $ \bar{N}N$. The potentials generated by these operators were calculated in Ref.~\cite{Fichet_2018}, and experimental bounds on (A) and (D) were placed in Ref.~\cite{Brax_2018}.  We write the generic nucleon-hidden sector coupling constant in terms of appropriate powers of a coupling scale $\Lambda$.

Note that for this effective description to be valid we require that the UV-completion of each interaction lies beyond the energy scale being probed by the experiment, which, in this case, does not typically exceed the $\mathcal{O}(\text{MeV})$ level.  This does not, however, imply that the effective theory description holds at accelerator energy scales, or even that it holds with equivalent validity at different fifth force experiments.  For instance, consider generating the interaction in (B) by integrating out an $\mathcal{O}(\text{keV})$ mass scalar with $\mathcal{O}(10^{-3})$ couplings to $\mathcal{O}_{SM}$ and $\Bar{\psi}\psi$.  This will generate $\Lambda \sim \mathcal{O}(\text{MeV})$. The effective theory description breaks down at the $\mathcal{O}(\text{keV})$ level, well before accelerator energy scales, and even at wavelengths longer than those probed by some fifth force experiments.  As a result, bounds on this type of model would have to be computed for this specific scenario with the mediating field included where necessary.  With this in mind, we will not attempt to determine high energy constraints on these couplings which, for certain UV-completions, may be very stringent and potentially invalidate large regions of parameter space.  Instead we emphasise that it is necessary to remember throughout that the validity of the EFT description of the interactions may break down if the UV-completion enters at energies below those being probed experimentally.

A further UV consideration is that of naturalness.  The operators (A)-(C) explicitly break any symmetries that could have kept the dark sector states light.  As a result, depending on the magnitude of $\Lambda$ and the relevant UV-cutoff in the SM and/or dark sector, additional tuning or stabilisation mechanisms may be required.  Since this also depends on the microscopic physics, we will not attempt a discussion of any fine-tuning considerations, but emphasise that such aspects should be considered for complete models.

\begin{figure}[tt]
\begin{minipage}{0.45\textwidth}
\subfloat[]{
\begin{fmffile}{bubble3}
    \begin{fmfgraph}(50,50)
       \fmfleft{i}
       \fmfright{o}
       \fmfdot{v1,v2}
       \fmf{phantom,tension=5}{i,v1}
       \fmf{phantom,tension=5}{v2,o}
       \fmf{dashes,left,tension=0.4}{v1,v2,v1}
     \end{fmfgraph}
\end{fmffile}
}
\subfloat[]{
\begin{fmffile}{bubble4}
    \begin{fmfgraph}(50,50)
       \fmfleft{i}
       \fmfright{o}
       \fmfdot{v1,v2}
       \fmf{phantom,tension=5}{i,v1}
       \fmf{phantom,tension=5}{v2,o}
       \fmf{fermion,left,tension=0.4}{v1,v2,v1}
     \end{fmfgraph}
\end{fmffile}
}
\subfloat[]{
\begin{fmffile}{bubble5}
    \begin{fmfgraph}(50,50)
       \fmfleft{i}
       \fmfright{o}
       \fmfdot{v1,v2}
       \fmf{phantom,tension=5}{i,v1}
       \fmf{phantom,tension=5}{v2,o}
       \fmf{boson,left,tension=0.4}{v1,v2,v1}
     \end{fmfgraph}
\end{fmffile}
}
\vspace*{5mm}
    \caption{Feynman diagrams representing -$\Delta(q)$ for interactions (a) A and D (b) B and (c) C.}
    \label{fig:bubbles}
\end{minipage}
\hfill
\begin{minipage}{0.45\textwidth}
 \centering
\begin{fmffile}{diagram2}
\begin{fmfgraph*}(120,60)
   \fmfleftn{i}{1}
   \fmfrightn{o}{2}
     \fmf{phantom}{i1,v1}
     \fmf{fermion}{o1,v1}
      \marrow{d}{down}{bot}{$k_1$}{v1,o1}
      \marrows{f}{up}{top}{$q$}{i1,v1}
       \marrow{e}{up}{top}{$k_2$}{v1,o2}
\fmfdot{v1}
\fmf{fermion}{v1,o2}
\end{fmfgraph*}
\end{fmffile}
\vspace*{5mm}
    \caption{The cut of the bubble diagram in Fig~\ref{fig:bubbles}b.}
    \label{fig:my_label}
    \end{minipage}
\end{figure}

As the coupling is bilinear, the leading order contribution to the potential from these interactions is at the 1-loop level. In the notation of Sec.~\ref{GP}, the `effective' propagators, $\Delta(q)$, for these interactions are the negative of the bubble diagrams in Fig.~\ref{fig:bubbles}.  In order to make use of Eq.~\ref{potential}, we need the imaginary part of these loops. These are most simply computed via cuts according to the optical theorem for forward scattering:

\begin{equation}
\label{opticaltheorem}
    2\Im{\mathcal{M}(A \rightarrow{}A) } = \\ \sum_X \int d\Pi_X (2\pi)^4 \delta^4(p_A - p_X) \abs{\mathcal{M}(A \rightarrow{}X)}^2 ~~,
\end{equation}

where $d\Pi_X$ is the Lorentz invariant phase-space factor for the state $X$.  If $\mathcal{M}(A \rightarrow{}A)$ is a 1-loop amplitude then $\mathcal{M}(A \rightarrow{}X)$ corresponds to the tree-level cut, illustrated in Fig.~\ref{fig:my_label}. Noting the resemblance of the RHS of Eq.~\ref{opticaltheorem} to the structure of usual cross-section and decay rate calculations, it is natural to interpret this as the 4-momentum conserving decay of an (external) source of momentum $q$, to a final state of two equal mass hidden sector particles.   The form of $\rho(\mu^2)$ and corresponding potentials calculated in this way are given in Table~\ref{results}. Some additional details on the calculation of $\rho$ can be found in Appendix~\ref{appendix}.  We note that the potentials extracted here agree with those found in  Ref.~\cite{Fichet_2018}, with the exception of (C), for which we find a different functional form. The potentials are all attractive, as expected for scalar operator exchange, and behave as  $r^{-3}$, $r^{-5}$, $r^{-7}$ and $r^{-7}$ respectively in the short-distance limit.

\begin{table*}[h]
\begin{center}
\resizebox{\textwidth}{!}{\begin{tabular}{c |c|c}
\multicolumn{1}{c|}{Operator}& \multicolumn{1}{c|}{$\rho(\mu^2)$}& 
\multicolumn{1}{c}{$V(r)$} \\
\hline
(A) & $\frac{\eta}{8\pi^2}\left(1-\frac{4m^2}{\mu^2}\right)^{\frac{1}{2}}~\Theta(\mu^2 - 4m^2)$&$-\frac{\eta m}{8 \Lambda^2 \pi^3 r^2}K_{1}(2mr)$\\[15pt]

(B) &$\frac{\mu^2 \eta}{4\pi^2}\left(1-\frac{4m^2}{\mu^2}\right)^{\frac{3}{2}}~\Theta(\mu^2 - 4m^2)$&$-\frac{3 \eta m^2}{2\Lambda^4 \pi^3r^3}K_{2}(2mr)$\\[15pt]

(C) & $\frac{\mu^4 \eta}{32m^4\pi^2}\left(1 + \frac{12m^4}{\mu^4} - \frac{4m^2}{\mu^2}\right)\left(1-\frac{4m^2}{\mu^2}\right)^{\frac{1}{2}}~\Theta(\mu^2 - 4m^2)$ & $-\frac{3 m^3 \eta(5+m^2 r^2)}{8\Lambda^6 \pi^3 r^4}K_3 (2mr)$\\[15pt]

(D)&$\frac{\mu^4 \eta}{32\pi^2}\left(1 - \frac{4m^2}{\mu^2} + \frac{4m^4}{\mu^4}\right)\left(1 - \frac{4m^2}{\mu^2}\right)^{\frac{1}{2}}~\Theta(\mu^2 - 4m^2)$&\makecell{$-\frac{ \eta}{8\Lambda^6 \pi^3}\left(\frac{15m^3}{r^4}+\frac{m^5}{r^2}\right)K_1(2mr)$ - \\$ \frac{ \eta}{4\Lambda^6 \pi^3}\left(\frac{15m^2}{r^5}+\frac{3m^4}{r^3}\right)K_2(2mr)$}\\[15pt]
\hline
\end{tabular}}
\caption{ $\rho(\mu^2)$ and $V(r)$ for the interactions in Eq.~\ref{boxed} as computed via the optical theorem. $K_n$ is the $n^{\textnormal{th}}$ modified Bessel function of the second kind, $\Theta$ is the Heaviside step function, and $\eta $ takes a value of 1 if the field is self conjugate and $1/2$ if not.}
\label{results}
\end{center} 
\vspace{-5mm}
\end{table*}

\subsection{Multiple-Loop Exchange}
To further highlight the utility of this approach we consider the following generalisation of interaction (A):
\begin{equation}
    \frac{1}{\Lambda^{n-1}} \mathcal{O}_{SM}\phi^n ~~, 
\end{equation}
where $n$ is a positive integer and $\phi$ is explicitly real. At lowest order, SM-SM scattering proceeds via ($n$ - 1)-loop exchange.  The Feynman diagram in the case that $n=3$ is shown in Fig.~\ref{fig:higherbubbles}.  The cut diagram, as required to use the optical theorem, is a tree-level process, as illustrated in Fig.~\ref{fig:highercut}, and can be interpreted as the decay of an external momentum source $q$ into a final state $X$ consisting of $n$ equal mass particles of momentum $k_i$, for $i \in \{1,...,n\}$. The amplitudes for such processes are constants, leading to spectral densities  

\begin{equation}
\rho_n(\mu^2) = \frac{n!}{2\pi} I_n(\mu) ~~,
\end{equation}
that are proportional to the integral over the Lorentz-invariant phase-space of the final $n$-body state\footnote{Note that the factor of 1/$(n!)$ needed to account for the fact that the final state particles are identical is not included in $I_n$ as defined here but is instead absorbed into the prefactor. See Appendix \ref{higher}  for further details.} 
\begin{equation}
    I_n(\mu) = \int d\Pi_X  ~ (2\pi)^4 \delta^4\left(\mu - \sum_i^n k_i \right)~~.
\end{equation}
We note that $\mu$ must exceed the sum of the masses of the final state particles: $nm$.  Each $I_n$ is thus implicitly accompanied by a Heaviside step function $\Theta(\mu^2 - n^2m^2)$.  The $I_n$ can be computed recursively using the relations given in Appendix~\ref{higher}. For the case $n=3$ we find that
 \begin{equation}
     \rho_3(\mu^2) = \frac{3\sqrt{(\mu - m)(\mu + 3m)}}{128\mu^2 \pi^4}\left(\frac{(\mu - m)(\mu^2 + 3m^2)}{2} E(\tilde{k}) - 4m^2 \mu K(\tilde{k})\right) \Theta(\mu^2 - 9m^2) ~~,
 \end{equation}
where $K$ and $E$ are the complete elliptic integrals of the first and second kind and 
\begin{equation}\label{ktilde}
    \tilde{k} = \sqrt{\frac{(\mu + m)^3(\mu - 3m)}{(\mu - m)^3(\mu + 3m)}} ~~.
\end{equation}  
To obtain the corresponding potential we use numerical integration. Fig.~\ref{abc} shows the potential $V(r)$ for the case where the hidden scalars each have a mass of 1 eV. At short distances, the potential scales as $r^{-5}$, as dictated by dimensional analysis.  

\begin{figure}[t]
\begin{minipage}{0.45\textwidth}
\centering
\begin{fmffile}{bubble6}
 \vspace*{5mm}
    \begin{fmfgraph}(120,60)
   
       \fmfleft{i}
       \fmfright{o}
       \fmfdot{v1,v2}
       \fmf{phantom,tension=5}{i,v1}
       \fmf{phantom,tension=5}{v2,o}
       \fmf{dashes,left,tension=0.4}{v1,v2,v1}
       \fmf{dashes}{v1,v2}
            \end{fmfgraph}
\end{fmffile}

\vspace*{5mm}
    \caption{Feynman diagrams representing -$\Delta(q)$ for lowest order SM-SM scattering generated by the operator $\frac{1}{\Lambda^{2}} \mathcal{O}_{SM}\phi^3$.}
    \label{fig:higherbubbles}
\end{minipage}
\hfill
\begin{minipage}{0.45\textwidth}
 \centering
\begin{fmffile}{diagram3}
\begin{fmfgraph*}(120,60)
   \fmfleftn{i}{1}
   \fmfrightn{o}{3}
     \fmf{phantom}{i1,v1}
     \fmf{dashes}{o1,v1}
      \marrow{d}{down}{bot}{$k_1$}{v1,o1}
      \marrows{f}{up}{top}{$q$}{i1,v1}
       \marrowss{e}{up}{top}{$k_2$}{v1,o2}
        \marrow{g}{up}{top}{$k_3$}{v1,o3}
\fmfdot{v1}
\fmf{dashes}{v1,o2}
\fmf{dashes}{v1,o3}
\end{fmfgraph*}
\end{fmffile}
    \caption{The cut of the bubble diagram in Fig~\ref{fig:higherbubbles}.}
    \label{fig:highercut}
    \end{minipage}
\end{figure}

\subsection{Beyond QFT}\label{beyond}
\begin{figure}[t]
\begin{minipage}{0.5\textwidth}
\includegraphics[width=80mm]{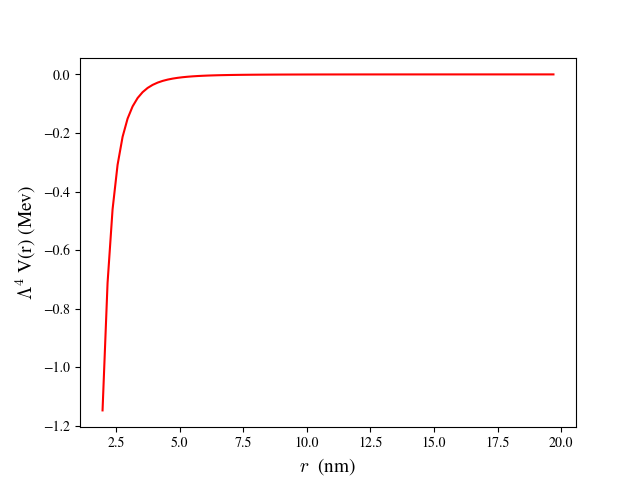}

\caption{The potential $V(r)$ generated by the operator $\frac{1}{\Lambda^{2}} \mathcal{O}_{SM}\phi^3$ for $m$ = 1 eV. }
\label{abc}
\end{minipage}
\hspace{5mm}
\begin{minipage}{0.38\textwidth}
\begin{tikzpicture}

\draw [lightgray,dashed]  (2,2.7)  -- (2,-2.7);
\draw [lightgray,dashed]  (1.2,2.7)  -- (1.2,-2.7);
\draw [lightgray,dashed]  (3.6,2.7)  -- (3.6,-2.7);
\draw [fill=cyan] (1.2,0) rectangle (2,1.8);
\draw [fill=cyan] (2,0) rectangle (3.6,-1.8);
\draw [<-]  (0,3) node [left] {$\rho(\mu^2)$}  -- (0,0) --(0,-3) ;
\draw [<-]  (5,0) node [below] {$\mu^2$} -- (0,0) node[left]{0};
\draw [very thick,black] (1.2,0)  -- (1.2,1.8)  -- (2,1.8)
       -- (2,0) ;
\draw [ very thick,black] (2,0)  -- (2,-1.8)  -- (3.6,-1.8)
       -- (3.6,0)  ;

\node[left] at (0,-1.8) {-1};
\node[above] at (2,2.7) {$\mu_0^2$};
\node[below] at (1.2,-2.7) {$\mu_0^2 - a\mu_0^2$};
\node[below] at (3.6,-2.7) {$\mu_0^2 + b\mu_0^2$};
\node[left] at (0,1.8) {1};

\end{tikzpicture}

\caption{Sketch of the form of $\rho_{v}(\mu^2)$ as defined in Eq.~\ref{violate}. }
\label{fig:sd}
\end{minipage}

\end{figure}

Given that it encodes, under a reasonable set of assumptions, fundamental principles such as unitary, causality, locality etc.\ it is tempting to consider what happens if the constraint  $\rho \geq  0$ is relaxed. As a crude first step in this direction we may consider, solely for illustrative purposes, a spectral density function of the form 
\begin{equation}
\label{violate}
\rho_v(\mu^2) = 
\Theta(\mu^2 - \mu_0^2 + a\mu_0^2)\Theta(\mu_0^2 - \mu^2) - \Theta(\mu^2 - \mu_0^2)\Theta(\mu_0^2+b\mu_0^2 - \mu^2) ~~,
\end{equation}
for constants $0 < a < 1$ and $b > 0$ which set the respective widths of the positive and negative regions as illustrated in figure \ref{fig:sd}.  The corresponding potential generated is

\begin{equation}
    V(r) = - \lambda^2 \frac{f(a) + f(-b)- 2e^{-\mu_0 r}\left(1+\mu_0 r\right)}{2\pi r^3} ~~,
\end{equation}
where
\begin{equation}
    f(x) = e^{-\mu_0r\sqrt{(1-x)}}\left(1+\mu_0r\sqrt{(1-x)}\right) ~~,
\end{equation}
and $\lambda$ denotes the nucleon-hidden sector coupling constant.

If $a = b$ the potential becomes  `screened', tending to a finite value in the short-distance limit. More interestingly, we note that for $b>a$, the potential develops a turning point. It is clear from Eq.~\ref{potential}, that this could not happen if $\rho \geq 0$ everywhere, demonstrating that turning points in the potential are a possible probe for the presence of different spin operators or, more speculatively, under the assumption that the interaction is generated by scalar operator exchange, violations of some fundamental postulates.

\section{Experimental Searches}\label{ES}
 In this section we review several of the main experimental approaches to searching for spin-independent fifth forces. Extensively using the mappings from inter-nucleon potentials presented in Ref.~\cite{Brax_2018}, we recast the  primary observables in terms of $\rho(\mu^2)$.  We then use these in conjunction with recent experimental results to place exclusion bounds on the models considered above, and investigate how the landscape of constraints varies for different forms of $\rho(\mu^2)$.

\subsection{Molecular Spectroscopy}
Simple molecular systems, for which both precision measurements and theoretical predictions of the energy spectra exist, can be used to probe anomalous forces with ranges on the order of the inter-nuclear separation \cite{Salumbides_2015,Salumbides:2013aga}.    `Exotic' systems, such as the antiprotonic helium ion, $\bar{p}^4$He$^{+}$, and muonic molecular deuterium ion, $dd\mu^{+}$, in which an electron is replaced by a heavier particle ($\bar{p}$ and $\mu^{-}$ respectively) are of particular interest, as the reduction in inter-nuclear separation affords the means to probe shorter-range forces.

To first order in perturbation theory, the presence of an additional inter-nucleon interaction with potential $V(r)$ would induce a shift to the energy of a given molecular state $\psi$ by an amount \begin{equation}
    \Delta E_{\psi} = \int d^3 \boldsymbol{r} ~ \psi^*(r) V(r) \psi (r) ~~.
\end{equation}
In terms of the spectral density function this becomes \begin{equation}
   \Delta E_{\psi} =  -\frac{\lambda ^2}{2\pi} \int d^3 \boldsymbol{r} ~ \psi^*(r) \frac{1}{r} \left( \int_0^{\infty}  d\mu ~ \mu ~ \rho(\mu^2) e^{-\mu r} \right)\psi (r) ~~.
\end{equation}

In practice, it is typically the energy to transition between a pair of states $\psi_1$ and $\psi_2$ that is measured experimentally. The modification to such an observable is thus $\Delta E_{\psi_1} - \Delta E_{\psi_2} \equiv \Delta E $. If the experimentally measured energy of a transition is consistent with its theoretical prediction from QED, bounds on anomalous forces can be obtained by demanding that the combined theoretical and experimental uncertainty, $\delta E = \sqrt{\delta E_{\textnormal{th}}^2 + \delta E_{\textnormal{exp}}^2} $, exceeds the anomalous shift $\Delta E$  \cite{Brax_2018}.

We perform this procedure for the transitions between the $(\nu =1, J=0) - (\nu =0, J=0)$  states of molecular hydrogen, $H_2$; the $(\nu =4, J=3) - (\nu =0, J=2)$ states of the hydrogen-deuterium ion, HD$^{+}$; the $(n =33, l=32) - (n =31, l=30)$ states of $\bar{p}^4$He$^{+}$; and the binding energy of the $(\nu =1, J=0)$ state of $dd\mu^{+}$. Here $\nu$, $J$, $n$ and $l$  denote the rotational, vibrational, principal and azimuthal quantum numbers of the state respectively. The integration over $r$ is completed numerically using the wave functions in Refs.~\cite{Salumbides:2013dua,Salumbides_2015}. The values of the combined theoretical and experimental uncertainty used to impose the bounds are quoted in Table~\ref{uncert}. In the small $r$ limit, the wave functions of all the states considered here, except that of $dd\mu^{+}$, are constant, and assume such small values that they can be neglected. In contrast, the wave function of the ($\nu = 1, J=0$) state of $dd\mu^{+}$ scales as $\sim r$ \cite{Brax_2018}, such that the tail at small $r$ is non-negligible. We thus adopt the practice used in Ref. \cite{Brax_2018} of imposing a UV-cutoff in the integration at $r_{\textnormal{uv}} = 2$ fm, the approximate radius of the deuteron. 
\begin{table*}
\begin{center}
\begin{tabular}{ |c|c|c|c|c| } 
 \hline
  System & $H_2$ & $HD^{+}$ &  $\bar{p}^4$He$^{+}$ & $dd\mu^{+}$  \\ 
   \hline
  $\delta E$ & \hspace{0.1em} 3.9 neV  \cite{Salumbides:2013aga,Salumbides_2015} \hspace{0.1em}   & \hspace{0.1em}  0.33 neV  \cite{Salumbides:2013aga,Salumbides_2015} \hspace{0.1em} & \hspace{0.1em}  24.81 neV \cite{Salumbides:2013dua} \hspace{0.1em}   & \hspace{0.1em}  0.7 meV \cite{Salumbides:2013dua} \hspace{0.1em}  \\ 

 \hline
\end{tabular}
\caption{The values of the combined theoretical and experimental uncertainty, $\delta E $, for the molecular spectroscopy experiments used to constrain anomalous forces in this work. }
\label{uncert}
\end{center}
\end{table*}

  \subsection{Bouncing Neutrons}
  Another approach is to bounce `ultracold' neutrons from a plane mirror \cite{Brax_2011,Brax_2013,Pignol_2015,Nesvizhevsky_2004}. For sub $\sim$ 10 $\mu$m bouncing heights, the neutron vertical motion is quantized due to the interaction with the terrestrial gravitational field \cite{Nesvizhevsky_2004}. At a vertical distance $z$ above the mirror, the neutrons experience a gravitational potential of strength  $V(z) = m_ngz$, with $m_n$ the mass of the neutron, and $g$ the local gravitational acceleration. The wave functions of allowed states are Airy functions
  \begin{equation}
      \psi_k (z) = c_k \textnormal{Ai}\left(\frac{z}{z_0} - \epsilon_k \right) ~~,
  \end{equation}
  where $\epsilon_k$ denotes the sequence of negative zeros of the Airy function i.e. Ai($-\epsilon_k$) = 0;  $z_0 = \left(\hbar^2 / (2m_n^2g)\right)^{\frac{1}{3}}$; and $c_k$ is a normalisation constant \cite{Brax_2011}. The corresponding energies are $ E_k = m_ngz_0\epsilon_k$. In the presence of an anomalous inter-nucleon force, the neutron experiences an additional potential due to its interactions with nucleons in the mirror. Assuming the nucleon density in the mirror to be constant, the additional potential at a height $z$ is 
  \begin{align}
    \delta V(z) &= 
    2\pi \frac{\rho_{\textnormal{glass}}}{m_n}\int_{z}^{\infty} dz' \int_{0}^{\infty}  d\tilde{\rho} ~ \tilde{\rho} V(r) \\ &=  2\pi \frac{\rho_{\textnormal{glass}}}{m_n}\int_{z}^{\infty} dr ~ r(r-z) V(r) ~~, 
\end{align}
where $V(r)$ is potential between the neutron and a single nucleon at a distance $r = \sqrt{z'^2+\tilde{\rho}^2}$. The integration is over the bulk of the mirror, which is taken to be semi-infinite in extent. In terms of the spectral density function this becomes
\begin{equation}
    \delta V(z) = \frac{-\lambda^2 \rho_{\textnormal{glass}}}{m_{n}}\int_0^{\infty} d\mu ~ \frac{\rho(\mu^2)e^{-\mu z}}{\mu}  ~~.
\end{equation}
To first order in perturbation theory, this induces a shift in the $k^{\textnormal{th}}$ energy level of
\begin{equation}
    \delta E_k = \int_0^{\infty} dz ~ |\psi_k (z)|^2 \delta V(z) ~~.
\end{equation}
 The energy difference $E_{3}-E_{1}$ has been measured by means of Gravity-Resonance-Spectroscopy at the Institut Laue Langevin \cite{Cronenberg:2015bol}. The value was found to be consistent with the theoretical prediction of the unperturbed spectrum. Bounds on anomalous forces can thus be placed by demanding that the theorised anomalous shift $\delta(E_{3}-E_{1}) = \delta E_{3} - \delta E_{1}$ is less than 10$^{-14}$ eV, the experimental precision of the measurement \cite{Cronenberg:2015bol}.

\subsection{Experiments with Effective Planar Geometry}
A further class of experiments designed to search for fifth forces are those which measure the force of attraction between two objects, typically either spherical or planar in shape, at sub-millimetre separations \cite{Decca_2007,Fischbach_2001,Chen_2016,Smullin:2005iv,Decca_2005}. For parallel plate configurations, it is straightforward to compute the anomalous inter-plate potential analytically by integrating the pairwise potential over the plate volumes. Where the overall geometry is not planar, as in sphere-plane or sphere-sphere cases, this is considerably more involved. If the object separation is small relative to their sizes however, such systems can be treated under the proximity force approximation (PFA). This connects the force  $F(s)$ between two objects of finite curvature at separation $s$, to the potential per unit area between two infinite parallel plates of the same composition and spacing, $V_{\textnormal{pp}}(s)$, via a constant of proportionality $R_{\textnormal{eff}}$ \cite{BordagMichael2009AitC}
\begin{equation}
    F(s) = 2 \pi R_{\textnormal{eff}} V_{\textnormal{pp}}(s) ~~.
\end{equation}
For plate-sphere configurations $R_{\textnormal{eff}}$ assumes the value of the sphere radius, and for 2 spheres of radii $R_1$ and $R_2$, $R_{\textnormal{eff}} = \sqrt{R_1 R_2}$.

In computing $V_{pp}$, it is important to account for the fact that the objects used are typically coated with several layers of different materials. Following Ref.~\cite{Brax_2018}, we characterise an object of bulk density $\sigma$ coated in $n$ successive layers of density $\sigma_n$ and thickness $\Delta_n$ by a piece-wise density profile, $\gamma(z)$:
\[\gamma(z) = 
\begin{cases}
\sigma_n & \hspace{2.5em} 0 < z < \Delta_n \\
\sigma_{n-1} & \hspace{2.5em} \Delta_n < z < \Delta_n + \Delta_{n-1} \\
\vdots\\
\sigma & \hspace{2.5em}  z > \sum_i^{n} \Delta_i
\end{cases}{} ~~.
\]
Denoting the anomalous  potential between a single pair of nucleons as $V(r)$, it follows that the potential per unit area $V_{pp}$ between plates with density profiles $\gamma_1(z)$, and $\gamma_2(z)$ separated by a distance $s$ is
\begin{equation}
    V_{\textnormal{pp}}(s) = \frac{2 \pi}{m_n^2} \int_{0}^{\infty} d\tilde{\rho} ~ \tilde{\rho} \int_{0}^{\infty} dz_1 \gamma_1(z_1) \int_{0}^{\infty} dz_2 \gamma_2(z_2) V(r) ~~,
\end{equation}{}
where $r = \sqrt{\tilde{\rho}^2 + (s + z_1 + z_2)^2}$.
In terms of the spectral function this becomes 
\begin{align}
    V_{\textnormal{pp}} &=  - \frac{ \lambda^2 }{m_n^2} \int_0^{\infty} d\mu ~ \frac{\rho(\mu^2)}{\mu^2} e^{-\mu s} J^{1}_{n_1}(\mu) J^{2}_{n_2}(\mu) ~~, 
\end{align}{} where \begin{equation}
    J^{i}_{n_i}(\mu) = \sigma_n + \sum_{l=1}^{n_i} \left(\sigma_{l-1} - \sigma_{l}\right) \textnormal{exp}\left(-\mu \sum_{i=l}^{n_i} \Delta_{i} \right) ~~.
\end{equation}\\
In this work we follow the treatment of Ref.~\cite{Fischbach_2001} in interpreting the results of force measurements between a polystyrene sphere of radius $R$ = 95.65 $\mu$m and a flat sapphire disk  made using Atomic Force Microscopy (AFM) by Mohideen et al.\ \cite{PhysRevA.62.052109}. Measurements were recorded for sphere-plate separations in the range 62 - 350 nm. The sphere and disk had bulk densities of $\rho_{\textnormal{pol}}$ = 1.06 gcm$^{-3}$ and $\rho_{\textnormal{sap}}$ = 3.98  gcm$^{-3}$ respectively, and were each coated in a $\Delta = 86.3$ nm layer of gold of density $\rho_{\textnormal{Au}}$ = 18.88 gcm$^{-3}$. According to the PFA, the anomalous force between the bodies when at a separation $s$ is
\begin{equation}
     F(s) = - \frac{2\pi R \lambda^2 }{m_n^2} \int_0^{\infty}  d\mu ~ \frac{\rho(\mu^2)e^{-\mu s}}{\mu^2}  \left(\rho_{\textnormal{Au}}+\left(\rho_{\textnormal{sap}}-\rho_{\textnormal{Au}}\right)e^{-\mu \Delta}\right)\left(\rho_{\textnormal{Au}}+\left(\rho_{\textnormal{pol}}-\rho_{\textnormal{Au}}\right)e^{-\mu \Delta}\right) ~~.
\end{equation}
From this expression it is clear that the largest forces and hence strongest constraints will come from the smallest value of $s$.  The rms deviation of the measured force $F_{\textnormal{exp}}$, from the theoretical prediction of the sum of established (Casimir and electrostatic) forces, $F_{\textnormal{th}}$, for the $N$ = 2583  measurements made across all separations,
\begin{equation}
\sigma_{\textnormal{rms}} = \sqrt{\frac{\sum\left(F_{\textnormal{exp}}-F_{\textnormal{th}}\right)^2} {N} } ~~,
\end{equation}
was determined to be 3.8 pN. As this exceeds the quoted experimental uncertainty on the measurement at $s =$ 62 nm of 3.5 pN, we cast (conservative) bounds on anomalous forces by demanding that $|F(\textnormal{s = 62 nm})| \leq \sigma_{\textnormal{rms}}$.

 \subsection{Cold Neutron Scattering}
 The scattering of non-relativistic neutrons from nuclei can be used to constrain anomalous forces \cite{Nesvizhevsky_2008,Frank_2004,Kamiya_2015}. In the SM, low-energy neutron scattering can be treated as a 4-fermion interaction with a scattering amplitude, $f(\boldsymbol{q})$,  that is both isotropic and independent of the 3-momentum transfer, $\boldsymbol{q}$ \cite{Nesvizhevsky_2008}. New physics could, in general, generate both contact and long-ranged contributions to $f(\boldsymbol{q})$. Scattering is typically described by introduction of a coherent scattering length, $l(\boldsymbol{q})$, which we define as
 \begin{equation}
     l(\boldsymbol{q}) =   l_{\textnormal{SM}}  + l_{\textnormal{BSM}}(\boldsymbol{q}) = - f(\boldsymbol{q}) = \sqrt{\frac{\sigma(\boldsymbol{q})}{4\pi}} ~~.
\end{equation} 
Nuclear scattering lengths have been measured for nearly all stable nuclei, using a variety of experimental methods \cite{KOESTER199165}. These can be broadly classed into two categories according to whether they measure at zero or non-zero scattering angle. Whilst the former, collectively referred to as ``optical'' techniques, are only sensitive to contact interactions, those in the latter measure both contact and non-contact contributions. Such techniques include Bragg diffraction, which measures at $q = q_{\textnormal{BD}}$, and the transmission method, which measures the total cross-section over all angles for high energy neutrons of momentum $k_{\textnormal{in}} \sim $ 1 eV. The scattering length extracted in this case is an angular average
\begin{equation}
    \Bar{l}_{\textnormal{tr}}(k_{\textnormal{in}})= \frac{1}{2}\int_{0}^{\pi} d\theta ~ \sin(\theta) ~ l\left(4k_{\textnormal{in}}^2 \sin^{2}\left(\theta /2\right)\right) ~~.
\end{equation} 
The constructions $l_{\textnormal{opt}}(0) - l_{\textnormal{Bragg}}(q_{BD})$ and   $l_{\textnormal{opt}}(0) - \Bar{l}_{\textnormal{tr}}(k_{\textnormal{in}})$, eliminate the contact terms and thus provide a direct measure of the long-ranged contributions to  $l_{\textnormal{BSM}}$.  By the Born approximation, $l_{\textnormal{BSM}}$ is related to the momentum space scattering potential, $V(\boldsymbol{q})$, via
\begin{equation}
    l_{\textnormal{BSM}}(\textbf{q}) = 2m_N V(\textbf{q}) = -2m_N\lambda^2 \int_{0}^{\infty} d\mu^2 \frac{   \rho(\mu^2)}{\abs{\textbf{q}}^2+\mu^2} ~~. 
\end{equation}
Note that for the toy loop exchange interactions considered above, $V(\textbf{q})$ is formally infinite. It can be regularised by making an appropriate subtraction (see \cite{Weinberg:1995mt}), with the penalty that the result becomes dependent on the arbitrarily chosen subtraction point.  Since this subtraction implies sensitivity to details of the UV and hence model dependence, we will not include these limits in the exclusion plots to follow.

 \subsection{Lunar Perihelion Precession}
 On an astrophysical scale, fifth forces can be probed by a search for anomalous orbital precession \cite{Adelberger_2003}. For the Moon, high precision measurements of this have been made by means of lunar laser ranging \cite{Merkowitz:2010kka}. Deviations of the measured precession from that expected from understood sources, such as the Earth's quadrupole moment, could be indicative of a fifth force. Denoting the anomalous potential between a pair of nucleons of mass $m_n$ at a separation $r$ to be $V(r)$, the macroscopic potential between the Earth and Moon is $\frac{M_M M_E}{m_n^2}V(r)$. The additional force in the radial direction is thus $F(r) = -\frac{M_M M_E}{m_n^2}\partial_r V(r)$. Assuming this to be considerably smaller than the gravitational attraction, it can be treated as a perturbation to the Earth-Moon orbit equation, which at second order modifies the orbital frequency. The precession angle between two successive perihelions can be shown to be 
 \begin{equation}
     \delta \theta = \left. - \frac{\pi a^3}{Gm_n^2 (1 - \epsilon^2)}\partial^2_{r} V(r) \right|_{r = a} ~~,
 \end{equation}
 where $a$ and $\epsilon$ are the semi-major axis and eccentricity respectively of the unperturbed orbit \cite{Brax_2018}. Recasting this in terms of the spectral density function yields
\begin{equation}
    \delta \theta =   \frac{\lambda^2}{Gm_n^2(1 - \epsilon^2)}\int_{0}^{\infty}  d\mu ~ \mu ~ \rho(\mu^2) e^{-\mu a} \left[1 + \mu a + \frac{(\mu a)^2}{2}\right]  ~~.
\end{equation}
Lunar laser ranging measurements give
\begin{equation}
    \delta \theta < 10^{-10} ~~,
\end{equation}
which can be used to set limits on anomalous forces \cite{Brax_2018}. 

\section{The Fifth Force Landscape}\label{landscape}
\begin{figure}
\centering
\begin{tabular}{P{0.5\textwidth}P{0.5\textwidth}}
  \includegraphics[width=70mm]{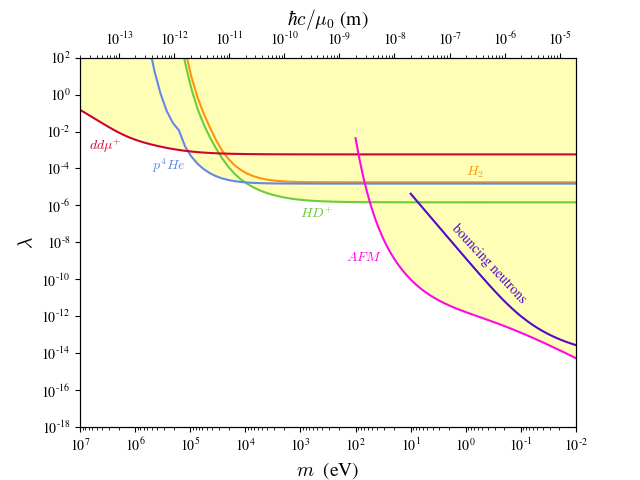}
 &   \includegraphics[width=70mm]{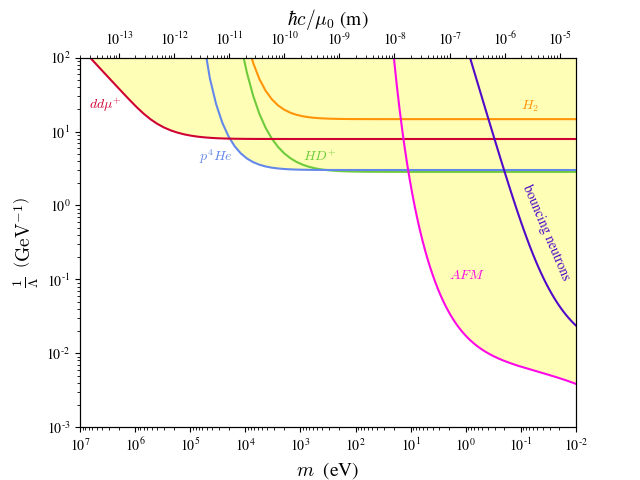} \\[6pt]
(i)  Yukawa interaction & (ii) Interaction (A) \\[15pt]
 \includegraphics[width=70mm]{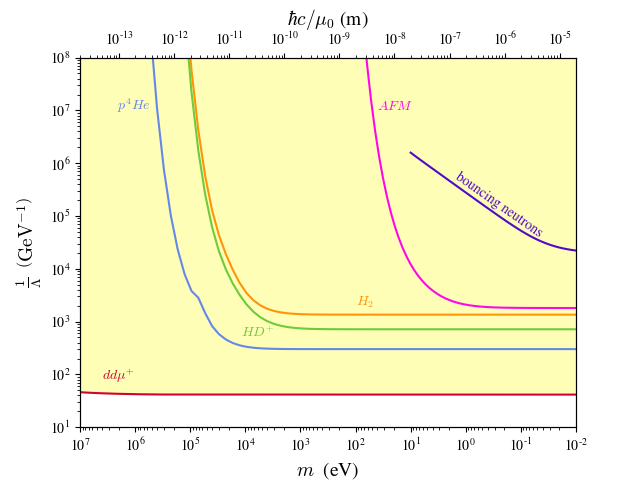} &   \includegraphics[width=70mm]{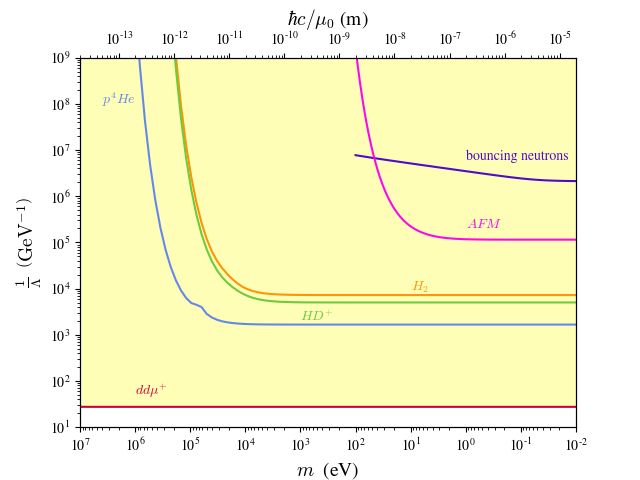} \\[6pt]
(iii) Interaction (B) & (iv) Interaction (C)\\[15pt]
 \includegraphics[width=70mm]{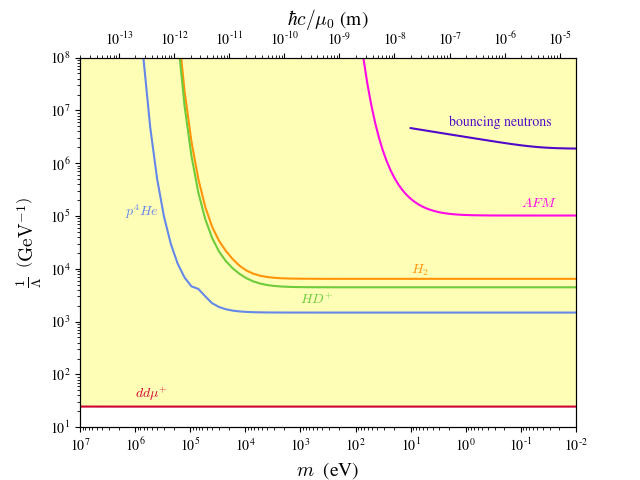} &   \includegraphics[width=70mm]{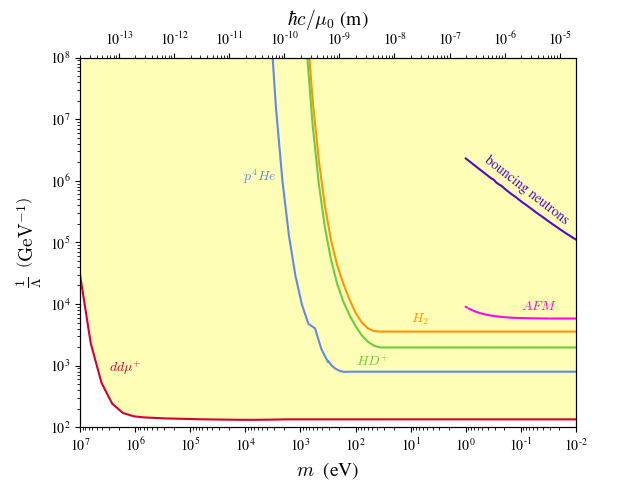} \\[6pt]
(v) Interaction (D)  & (vi) $\frac{1}{\Lambda^{2}} \mathcal{O}_{SM}\phi^3 \equiv$ Interaction (E) \\[5pt]
\end{tabular}
\caption{Experimental bounds on the parameter spaces of the various hidden sector models considered in this work. All dark scalar fields are assumed to be real. The shaded regions are excluded at the 95\% CL.}
\label{graphs}
\end{figure}

The plots in figure~\ref{graphs} (i)-(v)  show the results of applying the above limit-setting procedures to the Yukawa potential and each of the 1-loop interactions considered in Table~\ref{results} in turn. We assume that the hidden scalar in interaction (A) and (D) is self-conjugate (real), whilst the dark fermion and dark vector in interactions (B) and (C) are not. Figure~\ref{graphs} (vi) shows the bounds on 2-loop real scalar exchange, which we shall henceforth refer to as interaction (E).  For both interaction (A) and the Yukawa interaction, the experimental reach improves for lighter states. The strongest bounds come from bouncing neutrons and planar experiments which probe the longest distance scales. Conversely, for interactions (B)-(E), the constraint landscapes are dominated by the bound from the shortest distance experiment: $dd\mu^{+}$. This reversal in constraint hierarchy arises from the short-distance behaviour of these potentials which scale as $r^{-5}$, $r^{-7}$, $r^{-7}$ and $r^{-5}$ respectively, parametrically favouring experiments sensitive to shorter distance scales. As a result, the bounds from these experiments gain weight in carving out the exclusion regions for these potentials compared to those of interaction (A) and the Yukawa interaction, which scale as  $r^{-3}$ and $r^{-1}$ respectively in the short-distance limit. Bounds from lunar laser ranging are subdominant to those from shorter-range experiments for interactions (A)-(E) and are thus not plotted.  Given their reliance on the form of the subtraction used, and hence the details of the UV physics, the bounds from cold neutron scattering are less robust than those from the other experimental probes where no such regularisation is required. We therefore omit these from our plots also. The limits for (A) and (D) match the published results in Ref.~\cite{Brax_2018}.

We further note that for all of the loop interactions considered here, the effective coupling scale is well below collider scales, and in some cases as low as the $\mathcal{O}(10\text{ MeV})$ scale.  As a result, the UV-completion of these operators must enter, and be phenomenologically relevant, before reaching collider energies, and possibly even below the scales probed by some of the shortest-distance fifth force experiments.  Whether or not the UV-completion is in tension with observations is model-dependent.  For instance, if a fifth force experiment is probing below the eV scale, then states with extremely small couplings, almost as light as this scale, could be responsible for UV-completing the interaction.  Such states may alter cosmological and astrophysical observables, as well as being potentially observable in intensity frontier experiments, such as beam dump, missing energy, or rare meson decay searches. They could also modify the limits from fifth force experiments that probe shorter distances.  We emphasise that, for any given UV-completion, all of these possibilities should be considered.  However, since it is beyond the scope of this paper to map out the landscape of possible UV-completions of the interactions themselves, we present the fifth force results in terms of the effective coupling scale alone and keep in mind that a low-scale UV-completion is required.

\begin{figure}[t]
    \centering
    \includegraphics[width=90mm]{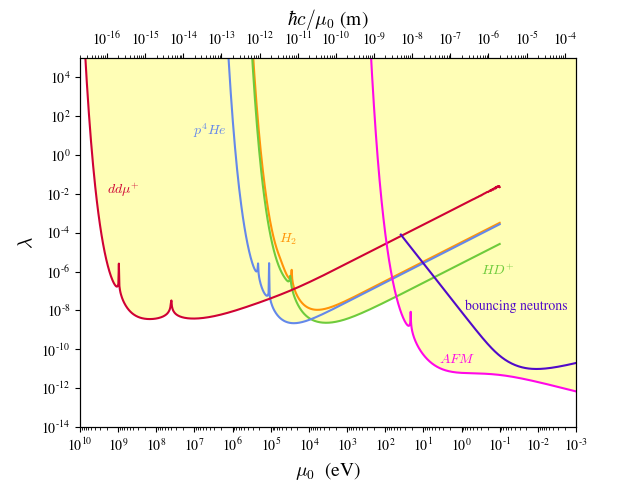}
    \caption{Bounds on the ($\mu_0$ , $\lambda$) plane for the interaction generated by $\rho_v$, as defined in Eq.~\ref{violate}, for the case $a = 0.1$, $b = 0.2$. The shaded regions are excluded at the 95\% CL.}
    \label{fig:viobounds} 
\end{figure}
Finally, let us turn to the constraint landscape for the spectral function $\rho_v$ defined in Eq.~\ref{violate} which, for scalar operator exchange, very crudely models the type of scenario that could arise if a fundamental principle such as unitarity, causality, or locality is violated.  For illustration, we take the values $a = 0.1$, $b = 0.2$. The width of the negative region is chosen to be twice that of the positive region such that potential exhibits a turning point in accordance with the analysis in Sec.~\ref{beyond}. The experimental bounds on this potential are shown in Fig.~\ref{fig:viobounds}. We note that in contrast to the spectral functions considered previously, the constraints coming from both the bouncing neutron experiments and molecular spectroscopy become more stringent as the scale $\mu_0$ is increased, before decreasing again.  Being the experimental arrangement most similar to the simple picture of two interacting point-like particles, this behaviour is most clearly understood in the case of molecular spectroscopy. We consider the force between two nucleons, as obtained from the derivative of the nucleon-nucleon potential, at the average inter-nuclear distance. As $\mu_0$ is increased the relative sensitivity of the experiment to positive and negative norm states varies and the force increases in magnitude before decreasing again, eventually changing sign. The value of $\mu_0$ at which the maximum magnitude occurs corresponds to the gross turning points occurring in Fig.~\ref{fig:viobounds}. This behaviour is a manifestation of the fact that the experimental force as a function of distance has a turning point whose location depends on the scale $\mu_0$.  By differentiating Eq.~\ref{potential} it is clear that this cannot  occur if $\rho(\mu^2) \geq 0$. The picture is more nuanced in the case of the AFM and bouncing neutron experiments as it is the force between extended objects which the experiments probe. When the inter-nucleon potential is summed over these objects one finds interference from different regions with varying magnitudes. 

These simple examples illustrate the breadth of possibilities for probing hidden sectors with fifth force experiments. Analyses assuming a simple Yukawa force are very well justified as a standard candle by which to begin exploring the hidden sector.  However, analysis techniques optimised specifically towards this form of force may miss more exotic possibilities for hidden sectors.  Indeed, we see that for generic loop-level exchange, corresponding to a continuum of states in $\rho(\mu^2)$, due to the higher dimension of the operator, the force becomes increasingly short-ranged and thus experiments probing the shortest distances begin to dominate sensitivity.  We have also briefly considered more exotic possibilities, such as non-monotonic force profiles which change sign over some distance scale. Discovery of an interaction of this kind could indicate the simultaneous presence of operators of different spin, for instance an attractive scalar channel which interferes with a repulsive vector channel, or some non-trivial role played by hidden gauge sector ghosts. Alternatively, such an observation could be the harbinger of something much more exotic, such  as very weakly coupled unitarity, causality, or locality violating sectors in nature.  Clearly, with such a rich and varied phenomenology possible, it is important that, much as at high energies, no stone is left unturned in the search for the influence of light weakly coupled hidden sectors.

\section{Conclusions} 
In recent years particle physics research has undergone somewhat of a phase transition, looking increasingly towards hidden sectors and the feebly interacting frontier.  When new hidden states mediate long-range forces one cannot appeal to decoupling and effective field theory to generalise their effects on low energy observables.  As a result, the historically popular approach has been to consider specific models.  In this work we have attempted to counter this model-dependent approach by showing that, under reasonable assumptions, the experimental effects of general scalar fifth forces may, using the K\"all\'en-Lehman representation, be captured by a single positive-definite function $\rho(\mu^2)$.

Using this prescription, accompanied by the optical theorem, a number of known results in the literature, obtained by other means, have been reproduced straightforwardly.  Furthermore, a number of commonly considered experimental observables have been recast in a general form, allowing a straightforward application to more complex hidden sectors.

Current experimental constraints reveal a hierarchy of sensitivity depending on the nature of the hidden sector operator, suggesting that comparison of experiments only for the assumption of a single scalar Yukawa force can be misleading since, in terms of the density of states, this  represents an extreme limit, $\rho(\mu^2) \propto \delta(\mu^2 - m^2)$, whereas much richer scenarios are possible.  Indeed, as expected, when more states participate in mediation the coupling dimension grows and the force becomes increasingly short-range, giving greater weight to experiments probing these scales.  This serves to illustrate that a variety of complementary searches are needed to fully explore hidden sector forces and echoes similar conclusions made in Ref.~\cite{Brax_2018}.

Finally, the breakdown of positivity of $\rho(\mu^2)$ could be the first tentative signal of something more interesting than simply a fifth force.  In this respect, the K\"all\'en-Lehman representation presents a rare non-perturbative tool with which to begin considering more speculative scenarios.  It is impossible to make completely general statements, however we have shown that in certain circumstances non-monotonic force profiles can occur.  This profile could imply the presence of different spin operators which interfere or, much more speculatively, under the assumption of scalar operator exchange, a departure from some fundamentals of QFT.  On this aspect we have barely scratched the surface.

\acknowledgments{
We are grateful to the Cambridge pheno working group for discussions.  This work has been  supported by STFC consolidated grants ST/P000681/1 and ST/S505316/1.
}

\appendix
\section{Calculation of $\rho$ for 1-Loop Interactions}\label{appendix}
For completeness, we provide further details on the computation of the spectral densities for the loop interactions (A)-(D) considered in this work. To find the imaginary part of  $\Delta(q)$ via the optical theorem, we seek the modulus squared of the cut amplitude, which we will henceforth denote as $\abs{\mathcal{M}(q \rightarrow{}X)}^2$. The Feynman diagrams for the \emph{negative}\footnote{This sign follows from our definition of $\Delta$ as a factor of the full scattering amplitude in Eq.~\ref{amp}.} of $\Delta(q)$ are shown in Fig.~\ref{fig:bubbles}. The required result for $\abs{\mathcal{M}(q \rightarrow{}X)}^2$ will thus be the \emph{negative} of the squared amplitude of the cut of these diagrams, as shown in Fig.~\ref{fig:my_label}. As noted above, it is natural to interpret such diagrams  as the decay of an external source of momentum $q$ into a final state $X$ consisting of a pair of equal mass hidden sector particles. In Table~\ref{matrix}, we present expressions for  $\abs{\mathcal{M}(q \rightarrow{}X)}^2$   for both the case where the hidden sector particles are identical (self-conjugate), and distinct (conjugate).  
\begin{table*}[h]
\begin{center}
\begin{tabular}{ |c|c|c| } 
\hline
 Model &  Identical & Distinct \\ 
  \hline
  (A) & -4 & -1  \\[4pt]
  (B) & $-8(q^2 - 4m^2)$ & -$2(q^2 - 4m^2)$ \\[4pt]
  (C) & $-\frac{q^4}{m^4}(1 + \frac{12m^4}{q^4} - \frac{4m^2}{q^2})$ &  $-\frac{q^4}{4m^4}(1 + \frac{12m^4}{q^4} - \frac{4m^2}{q^2})$ \\[4pt]
(D) & -$q^4(1 - \frac{4m^2}{q^2} + \frac{4m^4}{q^4})$& $-\frac{q^4}{4}(1 - \frac{4m^2}{q^2} + \frac{4m^4}{q^4})$ \\[4pt]
 \hline
\end{tabular}
\caption{$\abs{\mathcal{M}(q \rightarrow{}X)}^2$ where $\mathcal{M}(q \rightarrow{}X)$ corresponds to  the cut of $\Delta(q)$ for the case where the hidden particles in $X$ are identical (self-conjugate) and distinct (conjugate). (B) and (C) are the sum over all possible spin, and polarisation states, of the hidden sector particles in $X$, respectively. }
\label{matrix}
\end{center}
\end{table*}

For interactions (B) and (C), the given result is the sum over all possible spin and polarisation states respectively of the final state hidden sector particles. 
The factor of 4 difference between the self-conjugate and conjugate cases arises from the fact that we have not included an explicit symmetry factor in the couplings in our definition of these operators in Eq.~\ref{boxed}. 

As each of these results is independent of the momentum assigned to the final state particles $k_1$ and $k_2$, they can be factored out of the phase-space integral on the RHS of Eq.~\ref{opticaltheorem}.
The desired result is thus the product of the results in Table.~\ref{matrix}, with the integral over the Lorentz invariant phase-space of the final state:
\begin{equation}\label{i2}
    \int \frac{d^3k_1 d^3k_2}{(2\pi)^6 2k_1^0 2k_2^0}~(2\pi)^4 \delta^4(q - k_1 - k_2)  =  \frac{\kappa}{8\pi}\sqrt{1-\frac{4m^2}{q^2}} ~~,
\end{equation}
where $\kappa = 1/2 $ if the particles are identical (self conjugate), and 1 if distinct.  $\rho(\mu^2)$ then follows trivially as per Eq.~\ref{eq:rho}. 

\section{Calculation of $\rho$ for Exchange of $n$ Real Scalars}\label{higher}
The spectral density for $n-1$ scalar loop exchange, $\rho_n(\mu^2)$, can be calculated by an analogous procedure to that employed for the 1-loop case. The cut of  $\Delta(q)$ in this instance can be thought of as the decay of a momentum source $q$ to a final state $X$ consisting of $n$ hidden sector scalars each of mass $m$ and momentum $k_i$, $i \in \{1,...,n\}$.  The amplitude squared of this process is $-(n!)^2$. To obtain the RHS of the optical theorem we multiply this by the integral over the Lorentz invariant phase-space (LIPS) of the final state.  

For a generic state with $n$ \emph{distinct} particles of mass $m_i$ and 4-momentum $k_i$ the integral over LIPS is
\begin{equation}
    I_n(q;m_1,m_2,...,m_n) = \int \left(\prod_{i=1}^n \frac{d^{3}k_i}{(2\pi)^3 2k_i^0}\right) ~ (2\pi)^4 \delta^4(q - \sum_i^n k_i) ~~.
\end{equation}
This result should be implicitly associated with a $\Theta(q^2 - (\sum_i^n m_i)^2)$ function. We use the compact notation $I_n(q)$ to denote the specific case of equal masses. The required integral over the LIPS of the state $X$ in this notation is thus
\begin{equation*}
    \frac{I_n(q)}{n!} ~~,
\end{equation*}where the factor of $n!$  accounts for the fact that the final state particles are real and thus identical. 

Whilst such integrals are generally cumbersome, they can be evaluated recursively by performing just a single integral using the relation

\begin{equation}
    I_n(q;m_1,m_2,...,m_n) = \frac{1}{2\pi}\int ds ~ I_2(q; \sqrt{s},m_n) I_{n-1}(\sqrt{s}; m_1,...,m_{n-1}) ~~.
\end{equation}

The $\Theta$ functions associated with the $I_n$ restrict the integration over $s$ to the range  $\left(\sum_{i=1}^{n-1} m_i \right)^2$ to $(q - m_n)^2$ \cite{Davydychev_2004}.

We note that 
\begin{equation}
    I_2(q;m_1,m_2) \equiv \frac{1}{8\pi q^2}\sqrt{\lambda(q^2,m_1^2,m_2^2}) ~~,
\end{equation}
where 
\begin{equation}
    \lambda(x,y,z) = x^2 + y^2 + z^2 - 2xy - 2xz - 2yz ~~,
\end{equation}
is the generalisation of the result given in Eq.~\ref{i2}. 

For a state of 3 equal mass particles one obtains 
\begin{equation}
    I_3(q) = \frac{1}{128\pi^3 q^2}\int_{s_a}^{s_b} ds ~ \frac{ \sqrt{(s - s_a)(s-s_b)(s-s_c)}}{\sqrt{s}} ~~,
\end{equation}
where $s_a = 4m^2$, $s_b = (q-m)^2$ and $s_c = (q+m)^2$. 
This can be expressed in terms of the complete elliptic integrals of the first and second kind, $K$ and $E$ as
\begin{equation}
    I_3(q) = \frac{\sqrt{(q-m)(q+3m)}}{128\pi^3 q^2}\left(\frac{(q-m)(q^2 + 3m^2)E(\tilde{k})}{2} - 4m^2qK(\tilde{k})\right) ~~,
\end{equation} where $\tilde{k}$ is given by 
Eq.~\ref{ktilde}.

\bibliographystyle{JHEP}
\bibliography{biblio}

\end{document}